\RequirePackage[hyphens]{url}
\documentclass[twocolumn,prl,amsmath,superscriptadddress,longbibliography]{revtex4-1}
\usepackage{graphicx}
\usepackage[usenames,dvipsnames,svgnames,table]{xcolor}
\usepackage[colorlinks=true,linkcolor=RoyalBlue,citecolor=RoyalBlue]{hyperref}
\usepackage{soul}
\usepackage{bm}

\newcommand{\curl}{\nabla\times}

\newcommand{\dtot}[2]{\frac{d#1}{d#2}}
\newcommand{\bracket}[1]{\langle #1 \rangle}

\newcommand{\red}[1]{\textcolor{red}{#1}}

\begin{document}

\title{Thermodynamics of Energy Magnetization}

\author{Yinhan Zhang}

\affiliation{Department of Physics, Carnegie Mellon University,
  Pittsburgh, PA 15213, USA}

\author{Yang Gao}

\affiliation{Department of Physics, Carnegie Mellon University,
  Pittsburgh, PA 15213, USA} 

\author{Di Xiao}

\affiliation{Department of Physics, Carnegie Mellon University,
  Pittsburgh, PA 15213, USA}
  
\begin{abstract}
We construct the thermodynamics of energy magnetization in the presence of gravitomagnetic field.  We show that the free energy must be modified to account for the modification of the energy current operator in the presence of a confining potential.  The explicit expression of the energy magnetization is derived for a periodic system, and the Streda formula for the thermal Hall conductivity is rigorously established.  We demonstrate our theory of the energy magnetization and the Streda formula in a Chern insulator. 
\end{abstract}
  
\maketitle

\textit{Introduction}.---Recent years have seen a surge of interest in the thermal Hall effect, mainly due to its ability to probe charge neutral excitations in condensed matter systems with broken time-reversal symmetry~\cite{katsura2010a,onose2010,matsumoto2011a,hirschberger2015,banejee2018,kasahara2018,grissonnanche2019,samajdar2019}.  These systems have a vanishing (charge) magnetization, but they can still be characterized by an energy magnetization that arises from the circulating energy currents in thermodynamic equilibrium~\cite{cooper1997}.  It has been recognized that the energy magnetization plays an essential role in the theoretical understanding of thermoelectric transport~\cite{oji1985,cooper1997,xiao2006,qin2011,nomura2012,shitade2014,gromov2015}.  It must be properly discounted to obtain the correct transport coefficients and recover fundamental relations such as the Onsager relation and the Einstein relation~\cite{cooper1997, qin2011}.  In this context it is quite surprising that the theory of energy magnetization itself, and particularly its thermodynamics, remains in a primitive state.

The main challenge is the lack of a thermodynamic derivation of the energy magnetization. It has been identified that the conjugate force to the energy magnetization is the gravitomagnetic field~\cite{nomura2012,gromov2015,nakai2016,nakai2017}.  Therefore one should be able to obtain the energy magnetization as the derivative of the free energy to the latter.  However, these studies did not clarify what the exact expression of the free energy is and how the energy magnetization can be evaluated for a general extended system.  In fact, explicit calculations of the energy magnetization have referred to the existence of chiral edge states~\cite{nomura2012,nakai2016,nakai2017}.  Therefore it is not even clear whether the energy magnetization is truly a bulk quantity or not.  An expression for the energy magnetization has been previously derived~\cite{qin2011,shitade2014}, but the focus is on thermal transport using linear response theory, not on the thermodynamics of energy magnetization.

In this Letter we develop a theory to place the thermodynamics of the energy magnetization on a firm basis.  We first show that the bulk energy magnetization includes an anomalous contribution from the modified boundary energy current, which survives in the thermodynamic limit.  Consequently, the free energy should be modified to account for this anomalous contribution.  We then derive an explicit expression of the energy magnetization using the Maxwell relation, from which the Streda formula for the thermal Hall effect can be obtained~\cite{nomura2012}.  Finally, we demonstrate our theory of the energy magnetization and the Streda formula in a Chern insulator, and show that our theory is able to capture the contribution from the chiral edge states.

\begin{figure}[b]
\includegraphics[width=\columnwidth]{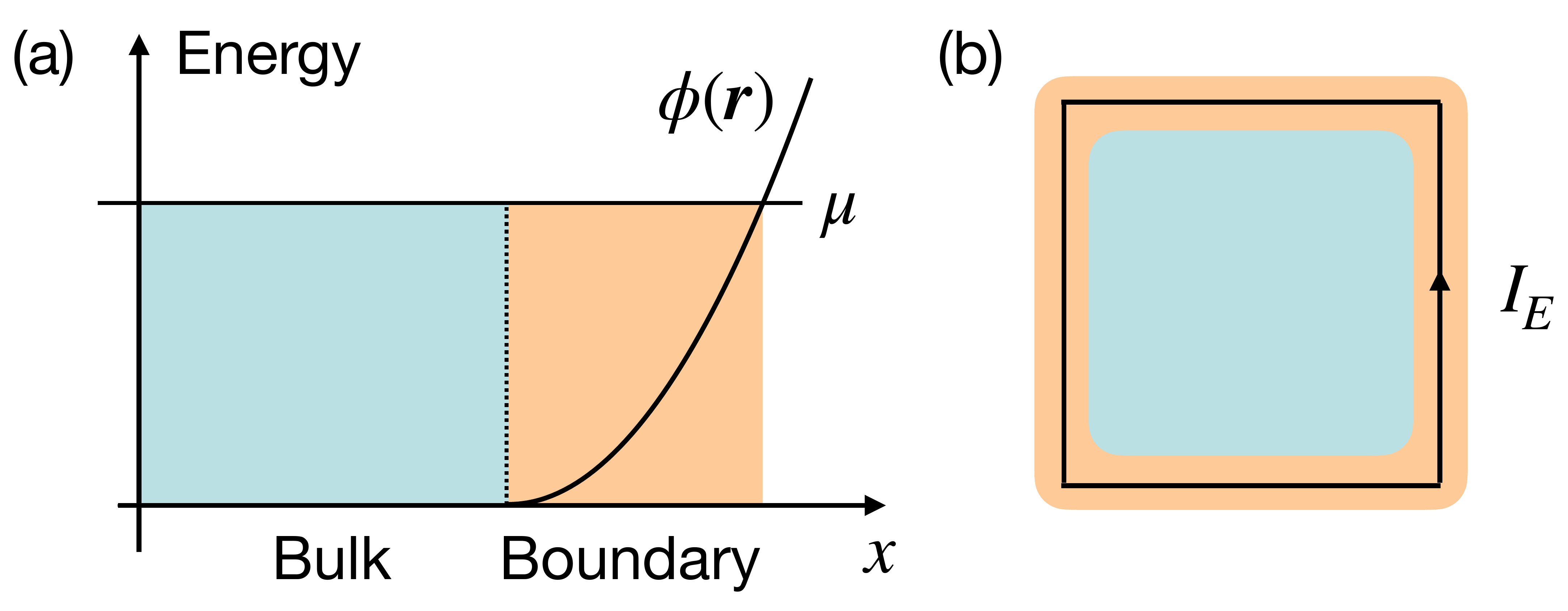}
\caption{\label{fig:boundary}(a) Side view: The confining potential $\phi(\bm r)$ separates the system into a bulk region and a boundary region. (b) Top view: The boundary energy current $I_E$ gives rise to an energy magnetization.}
\end{figure}

\textit{General considerations}.---The difficulty in deriving the energy magnetization is that it is only defined via the relation
\begin{equation} \label{def}
\bm j_E(\bm r) = \curl \bm M_E(\bm r) \;,
\end{equation}
where $\bm j_E$ is the energy current and $\bm M_E$ the energy magnetization.  Naively, one expects that $\bm M_E$ is given by the expectation value of $(1/2)\hat{\bm r} \times \hat{\bm j}_E$.  However, it is well known that for an extended system the current density alone is not sufficient to determine the corresponding magnetization~\cite{hirst1997,xiao2005,thonhauser2005,shi2007}. To circumvent this difficulty, our strategy is to first calculate the total magnetic moment of a finite system, then define the energy magnetization as the thermodynamic limit of the following expression~\footnote{One can replace $\bm j_E$ in Eq.~\eqref{mag} with Eq.~\eqref{def}.  The equation will only hold if the boundary term vanishes after an integration by parts.  For a finite system, this is always possible if we choose the boundary to be entirely outside the system.  For an extended system (with periodic boundary conditions), the integral can take arbitrary value depending on the choice of the boundary.}
\begin{equation} \label{mag}
\bm M_E = \lim_{\mathcal V \to \infty} \frac{1}{\mathcal V}
\int_{\mathcal V} d\bm r\, 
\frac{1}{2}\bm r \times \bm j_E \;,
\end{equation}
where $\mathcal V$ is the volume of the system.

Physically, a finite system can be realized by adding a confining potential $\phi(\bm r)$.  For simplicity, let us consider a two-dimensional system.  We assume that the confining potential is constant ($\phi_0$) inside the bulk and gradually increases to infinity as $r \to \infty$, as shown in Fig.~\ref{fig:boundary}(a).  We further assume that $\phi(\bm r)$ is sufficiently smooth such that a local chemical potential $\mu(\bm r) = \xi - \phi(\bm r)$ can be defined, where $\xi$ is the global chemical potential.  At this point, it is crucial to realize that $\phi(\bm r)$ not only changes the local chemical potential, but also modifies the energy current operator itself. This is because in addition to its internal energy, a particle also carries the potential energy $\phi(\bm r)$.  The energy current is thus given by~\cite{cooper1997,LandauV2}
\begin{equation} \label{modified_current}
\bm j_E^\phi(\bm r) = \bm j_E^0(\bm r) + \phi(\bm r) \bm j_N(\bm r) \;,
\end{equation}
where $\bm j_E^0(\bm r)$ is the expectation value of the energy current in the absence of the confining potential and $\bm j_N(\bm r)$ is the particle number current.  Since both $\bm j_E^0(\bm r)$ and $\bm j_N(\bm r)$ are equilibrium currents, we can express them in terms of their corresponding magnetizations and write $\bm j_E^\phi(\bm r)$ as
\begin{equation}\label{modified_current2}
\bm j_E^\phi(\bm r) = \curl \bm M_E^0(\bm r) + \phi(\bm r) \curl \bm M_N(\bm r) \;,
\end{equation}
where $\bm M_E^0$ is the magnetization of the unmodified energy current, and $\bm M_N$ is the particle number magnetization~\cite{xiao2005,thonhauser2005,xiao2006,shi2007}.  The $\bm r$-dependence of the magnetizations enters through the local chemical potential $\mu(\bm r)$, e.g., $\bm M_N(\bm r) = \bm M_N(\mu(\bm r)) = \bm M_N(\xi - \phi(\bm r))$.  \red{}

Since $\phi(\bm r)$ is a constant in the bulk, $\bm j_E^\phi(\bm r)$ is confined to the boundary area of the system.  Let us assume that the boundary is along the $y$ direction, then the boundary energy current is given by
\begin{equation} \label{boundary}
I_E = -\int_\text{bulk}^\infty dx\, \Bigl(\dtot{M_E^0}{x} + \phi(x) \dtot{M_N}{x}\Bigr) \;.
\end{equation}
It should give rise to a total energy magnetic moment, approximately $I_E\mathcal A$, where $\mathcal A$ is the area of the system [Fig.~\ref{fig:boundary}(b)].  The correction comes at the order of $\mathcal O(\sqrt{\mathcal A})$.  Therefore, in the thermodynamic limit the energy magnetization is simply given by $I_E$.  Integrating Eq.~\eqref{boundary} by parts and making use of the boundary condition that both $\bm M_E^0$ and $\bm M_N$ vanish as $r\to\infty$, we obtain
\begin{equation} \label{modified_mag}
\begin{split}
\bm M_E^\phi(\mu_0, T) &= \bm M_E^0(\mu_0, T) + \phi_0\bm M_{N}(\mu_0, T) \\
&\qquad + \int_{-\infty}^{\mu_0} d\lambda\, \bm M_N(\lambda, T) \;.
\end{split}
\end{equation}
where in the bulk the local chemical potential becomes a constant $\mu_0 \equiv \xi - \phi_0$.  One can verify that if we let $\phi_0$ depend on $\bm r$, then $\curl\bm M_E^\phi = \bm j_E^\phi$.

The energy magnetization $\bm M_E^\phi(\mu_0, T)$ in Eq.~\eqref{modified_mag} still depends on $\phi_0$, which can be arbitrary. It is more convenient to remove the $\phi_0$-dependence by introducing the heat magnetization $\bm M^{\phi}_{Q}\equiv \bm M^{\phi}_{E}-\xi \bm M_{N}$, which can be regarded as the energy magnetization defined with respect to the chemical potential. It is given by   
\begin{equation} \label{modified_hmag}
\begin{split}
\bm M^{\phi}_Q(\mu_0, T) &= \bm M_E^0(\mu_0, T) - \mu_0 \bm M_{N}(\mu_0, T) \\
&\qquad + \int_{-\infty}^{\mu_0} d\lambda\, \bm M_N(\lambda, T) \;.
\end{split}
\end{equation}
We see that the heat magnetization $\bm M^\phi_Q(\mu_0, T)$ only depends on the local chemical potential $\mu_0$ in the bulk.

The above analysis seems to suggest that the heat magnetization $\bm M^{\phi}_Q$ is a boundary-dependent quantity. Indeed, the superscript $\phi$ indicates that it is calculated using the energy current $\bm j_E^\phi$, which includes a modification due to the confining potential $\phi(\bm r)$ [see Eq.~\eqref{modified_current}]. However, the right-hand side of Eq.~\eqref{modified_hmag} makes no explicit reference to $\phi(\bm r)$.  Therefore, $\bm M^{\phi}_Q(\mu_0, T)$ should be regarded as a genuine bulk quantity and we will drop the superscript $\phi$ in the following.

\textit{The modified bulk free energy}.---Armed with the insight that a proper calculation of the energy magnetization must include the confining potential $\phi(\bm r)$, we now give a rigorous derivation. To evaluate the energy magnetization, we introduce an auxiliary vector field $\bm A_g$ that linearly couples to the energy current $\hat{\bm j}_E^\phi(\bm r)$ in the Hamiltonian (throughout this paper we have set $e=\hbar=k_{B}=1$)        
\begin{equation}\label{Ham}
\hat{H}^{\phi} = \int d\bm r\left(\hat{h}_0(\bm r)+\hat{n}(\bm r)\phi(\bm r)\right) - \int d\bm r\, \hat{\bm j}_E^\phi(\bm r) \cdot \bm A_g(\bm r) \;,
\end{equation}
where $\hat{h}_0(\bm r)$ is the hamiltonian density without external fields, $\hat n(\bm r)$ is the particle number density operator, and $\hat{\bm j}_E^\phi(\bm r) =\hat{\bm j}^{0}_{E}+\phi(\bm r)\hat{\bm j}_{N}(\bm r)$ is the energy current operator.  Here $\bm A_g$ can be regarded as a purely mathematical device: as we show below, the static response of the free energy to $\bm B_g = \curl\bm A_g$ yields the energy magnetization.  Physically, $\bm B_g$ is the gravitomagnetic field and $\bm A_g$ is its vector potential~\cite{gromov2015}.

Let us expand the free energy $\Omega^{\phi}_{A_g}$ with respect to $\bm A_g$ up to first order,
\begin{equation}\label{modified_free1}
\Omega^{\phi}_{A_g}\approx\Omega^{\phi}_0-\int d\bm r \bm A_g(\bm r)\cdot \bm j^{\phi}_E(\bm r) \;,
\end{equation}  
with $\Omega^{\phi}_0\equiv\Omega^{\phi}_{A_{g}=0}$. Here $\bm j^{\phi}_{E}(\bm r)$ is the statistical expectation of $\hat{\bm j}_E^\phi(\bm r)$ in the absence of $\bm A_g$, and it is exactly the current density appearing in Eq.~\eqref{modified_current}. Inserting Eq.~\eqref{modified_current2} into Eq.~\eqref{modified_free1}, and using partial integration, we find that the linear coupling term in Eq.~\eqref{modified_free1} becomes
\begin{equation}
-\int d\bm r \bm A_g(\bm r)\cdot \bm j^{\phi}_E(\bm r)
= -\int d\bm r \bm B_g(\bm r)\cdot \bm M^{\phi}_{E}(\bm r) \;.
\end{equation}
where $\bm M_{E}^{\phi}$ is given by Eq.~\eqref{modified_mag}.  We see that the conjugate force to $\bm M_E^\phi$ is indeed $\bm B_g$.

As discussed earlier, to get rid of the $\phi_0$-dependence of the energy magnetization, we can switch from the energy magnetization $\bm M^{\phi}_{E}$ to the heat magnetization $\bm M_{Q}$. By adding a term $-\xi \int d\bm r\, \bm B_g \cdot \bm M_{N}$  to Eq.~\eqref{modified_free1}, we finally obtain the free energy in terms of the heat magnetization $\bm M_Q$,
\begin{equation}\label{modified_free3}
\Omega_{b}(\mu_0, T, \bm B_g)=\Omega_0^{\phi}(\mu_0, T)- \int d\bm r \bm B_g(\bm r)\cdot \bm M_Q(\mu_0, T) \;.
\end{equation}
In the following we will simply write $\mu_0$ as $\mu$.

In Eq.~\eqref{modified_free3} the zero-field free energy $\Omega_0^\phi(\mu_0, T)$ still depends on $\phi$. It is convenient to introduce an effective bulk Hamiltonian density $\hat{\mathcal{H}}_{A}(\bm r)$ that makes no reference to $\phi$,
\begin{equation}\label{effH}
\hat{\mathcal{H}}_{A}(\bm r)=\hat{h}_0(\bm r) -\mu\hat{n}(\bm r)-\, \hat{\bm j}_Q(\bm r) \cdot \bm A_g(\bm r),
\end{equation}
where $\hat{\bm j}_Q(\bm r)\equiv \hat{\bm j}^{0}_{E}(\bm r)-\mu\hat{\bm j}_{N}(\bm r)$ is the heat current operator.  One can verify that the following free energy  
\begin{align}\label{modified_free4}
\Omega_b(\mu, T) = &-\frac{1}{\beta}\ln\text{Tr}[e^{-\beta \int d\bm r\hat{\mathcal{H}}_{A}(\bm r)}]\notag\\&-\int d\bm r \bm B_g(\bm r)\cdot \int_{-\infty}^\mu d\lambda \, \bm M_N(\lambda, T)
\end{align}
generates Eq.~\eqref{modified_free3} up to the first order in $\bm B_g$.  The first term is the standard expression, while the second term originates from the modification to the energy current by the confining potential.  In contrast, for particle number magnetization, such a modification does not exist and no extra term is needed in the free energy.  Equation~\eqref{modified_free4} is our central result.
 
\textit{Heat magnetization}.---It follows from Eq.~\eqref{modified_free3} that the zero-field heat magnetization is given by  
\begin{equation}
\bm M_{Q}=-\lim_{\bm B_g\rightarrow 0}\left(\frac{\partial\Omega_{b}}{\partial\bm B_g}\right)_{\mu, T} \;.
\end{equation}
However, it is more convenient to calculate $\partial \bm M_{Q}/\partial \mu$, or $\partial \bm M_{Q}/\partial T$ with the help of the Maxwell relations using the modified free energy $\Omega_{b}$,
\begin{align}
\left(\frac{\partial{\bm M_{Q}}}{\partial \mu}\right)_{\bm B_g, T}&=\left(\frac{\partial N}{\partial \bm B_g}\right)_{T, \mu} \;, \label{maxwell1} \\
\left(\frac{\partial{\bm M_{Q}}}{\partial T}\right)_{\bm B_g, \mu}&=\left(\frac{\partial S}{\partial \bm B_g}\right)_{T, \mu} \;, \label{maxwell2}
\end{align} 
where $N = -\partial\Omega_b/\partial \mu|_{\bm B_g, T}$ is the particle number and $S = -\partial\Omega_b/\partial T|_{\bm B_g, \mu}$ is the entropy.

In the following we sketch the derivation of the heat magnetization using Eq.~\eqref{maxwell1}.  The right-hand side of Eq.~\eqref{maxwell1} is the static linear response coefficient of the particle number $N$ to the external field $\bm B_g$ in equilibrium. Although the free energy is modified, we find that the particle number is still given by $N=\text{tr}[\hat{\rho}_A \hat{N}]$ with the statistical distribution $\hat{\rho}_A=\exp(-\beta\int d\bm r \hat{\mathcal{H}}_{A}(\bm r))/Z_{A}$~\footnote{This would not be the case had we not included the correction to the free energy.}. Therefore we can construct a static response function~\cite{kubobook1983} 
\begin{equation}
\bm \Pi_{n,j^{Q}}(\bm q)=- \lim_{\omega_{n}\rightarrow 0}\int d\tau e^{i\omega_n \tau}\bracket{T_{\tau}\hat{\rho}_0\hat{n}_{\bm q}(\tau)\hat{\bm j}^Q_{-\bm q}(0)} \;,
\end{equation}
where $\hat{\rho}_0=\hat{\rho}_{A=0}$, and $\hat{n}_{\bm q}$ and $\hat{\bm j}^Q_{\bm q}$ are the Fourier component of the particle number density $\hat{n}(\bm r)$ and the heat current density $\hat{\bm j}_{Q}(\bm r)$, respectively.  The density fluctuation induced by the probing field $\bm A_g(\bm q)$ is $\delta n_{\bm q}=\bm \Pi_{n,j^Q}(\bm q)\cdot\bm A_g(\bm q)$.  To extract the response to $\bm B_g$, we expand $\bm{\Pi}_{n,j^Q}(\bm q)$ up to first order in $\bm q$, and only retain the antisymmetric part after taking the $\bm q \to 0$ limit.  Using Eq.~\eqref{maxwell1}, we find 
\begin{equation}\label{densitymq}
\frac{\partial \bm M_Q}{\partial \mu}\bigg|_{\bm B_g\rightarrow 0}=\frac{i}{2}\lim_{\bm q\rightarrow 0}\nabla_{\bm q}\times \bm \Pi_{n,j^{Q}}(\bm q) \;.
\end{equation}
Integrating the above equation with respect to $\mu$ and using the boundary condition $\bm M_{Q}\rightarrow 0$ when $\mu\rightarrow -\infty$, we can obtain the heat magnetization. 

In the following we apply our theory to a noninteracting periodic fermion system.  The heat magnetization is given by~\cite{supp} 
\begin{equation}\label{HMag}
\bm M_{Q}= \int [d\bm q]\Bigl(\bm{m}^Q_{\bm q}f(\tilde{\varepsilon}_{\bm q})-\int^{\infty}_{\tilde{\varepsilon}_{\bm q}}d\lambda f(\lambda)\lambda\bm{\Omega}_{\bm{q}}\Bigr) \;,
\end{equation}
where $\int [d\bm q]$ is a shorthand for $\int d\bm q/(2\pi)^3$, $\tilde{\varepsilon}_{\bm q}\equiv\varepsilon_{\bm q}-\mu$, and $f(x)=1/(e^{\beta x}+1)$ is the Fermi-Dirac distribution.  Summation over the band index has been omitted.  As shown in the Supplementary Material~\cite{supp}, the heat magnetization has a very intuitive interpretation in terms of the wave packet of Bloch electrons.  The first term in Eq.~\eqref{HMag} contains the quantity $\bm m^Q_{\bm q}=i\bracket{\nabla_{\bm q} u|\times(\hat{\mathcal{K}}^{2}-\tilde{\varepsilon}^2_{\bm{q}})|\nabla_{\bm q}u}/4$ with $|u\rangle$ being the periodic part of the Bloch function, which can be interpreted as the heat magnetic moment due to the self-rotation of the wave packet.  The second term involves the Berry curvature $\bm \Omega_{\bm q}=\nabla_q\times\bracket{u|i\nabla_{\bm q}u}$, and can be understood as the global circulating motion of the center-of-mass of the wave packet~\cite{xiao2006,matsumoto2011a,zhang2016}.  Derivation using the Maxwell relation~\eqref{maxwell2} gives the same expression~\cite{supp}.

We have also verified that our heat magnetization indeed leads to the correct thermal Hall coefficient~\cite{supp}, which agrees with previous linear response calculation.~\cite{qin2011}.

\textit{Streda formula}.---With the thermodynamics of the energy magnetization firmly established, we now discuss the Streda formula for the thermal Hall effect, which was first proposed in Ref.~\cite{nomura2012}.  Previous calculations have made explicit reference to the chiral edge states~\cite{nomura2012,nakai2016,nakai2017}.  Here we give a thermodynamic derivation.  The key step is the realization that the gravitomagnetic field $\bm B_g$ introduced earlier satisfies Faraday's law, i.e., $\curl \bm E_g = \partial_t \bm B_g$, where $\bm E_g$ is the  gravitoelectric field.  This can be rigorously established using the Newton-Carton (NC) geometry with a temporal torsion~\footnote{The relation between the thermal Hall effect and the Newton-Carton (NC) geometry with torsion is discussed in Ref.~\cite{gromov2015}.  The temporal torsion is given by $T_{\mu\nu}=\partial_{\mu}n_{\nu}-\partial_{\nu}n_{\mu}$, where $n_{\nu} \equiv e^{0}_{\nu}$ is the time-component frame field in the NC geometry. Thus, $B_g=-\epsilon_{jk}\partial_{j}(e^{-\psi}n_{k})$ and $E_{g,j}=e^{-\psi}T_{0j}$ with $j=1,2$ are realized with $n_{\nu}=[e^{\psi},n_{1},n_{2}]$. When $n_{j}=0$ and $\psi\neq 0$, the gravitational electric field $\bm E_g$ is given by $\bm E_g=-\nabla\psi$, where $\psi$ is the gravitational potential mentioned by Luttinger~\cite{luttinger1964}. When $n_j\neq 0$ and $\psi=0$, the coframe field driving the heat current is exactly $\bm A_g=-\bm n$.  When $\psi$ is time independent, one can obtain Faraday's law for the gravitational electromagnetic fields,  $\nabla\times\bm E_g=-\partial_{t}\bm B_g$.}.  Now let us consider an adiabatic process in which a time-dependent flux of $\bm B_g$ generates a circulating $\bm E_g$ around the boundary of some region.  Since the process is adiabatic, the change of the entropy can be related to the heat current by the continuity equation $T\partial_tS = -\nabla \cdot \bm j_Q$.  As the thermal Hall conductivity $\kappa$ is defined via $\bm j_Q = T\kappa\hat z \times \bm E_g$, we find $\kappa = T(\partial_t S/\partial_t B_g)$.  If the energy spectrum is gapped, we can get rid of the time derivative in the adiabatic limit and arrive at  
\begin{equation}\label{streda}
\kappa_{xy}=\Bigl(\frac{\partial S}{\partial B_g}\Bigr)_{\mu, T}
=\Bigl(\frac{\partial M_Q}{\partial T}\Bigr)_{T, B_g}  \;,
\end{equation}
where in the last step we have used the Maxwell relation.

Let us insert the heat magnetization in Eq.~\eqref{HMag} into the Streda formula~\eqref{streda}.  We find
\begin{equation}\label{PMPT}
\frac{\partial{\bm M_{Q}}}{{\partial T}}=\int [d\bm{k}]\Bigl(\bm{m}^{Q}_{\bm{k}}\partial_{T}f_{\bm{k}}+\frac{1}{T}\int_{\tilde{\varepsilon}_{\bm{k}}}^{\infty}f^{\prime}(\lambda)\lambda^{2}d\lambda\Omega_{\bm{k}}\Bigr) \;.
\end{equation}
For a band insulator, the first term in Eq.~\eqref{PMPT} can be dropped at low temperatures.  The second term is exactly the thermal Hall conductivity obtained previously~\cite{qin2011}. Thus the Streda formula is only valid for gapped systems at low temperatures.

 \begin{figure}[t]
  \includegraphics[width=1.0\columnwidth]{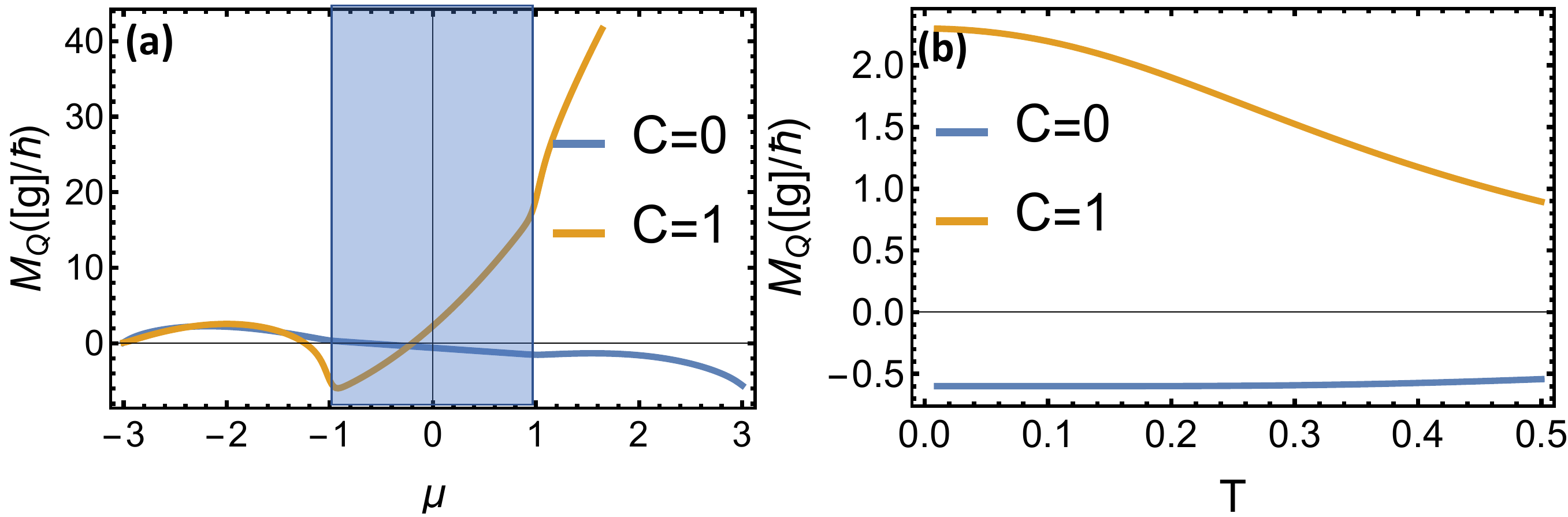} 
  \caption{The chemical potential and temperature dependences of the heat magnetization $\bm M_{Q}$: $\lambda=1$ and $\lambda=-3$ lead to the same gap region (blue region) as shown in (a), but different topological phases with $C=1$ and $C=0$. (a) the chemical potential dependence of $\bm M_{Q}$ at $T=0.03$; (b) the temperature dependence of $\bm M_{Q}$ at $\mu=0$.}
\label{fig:MQ}
\end{figure} 
       
\textit{Lattice model}.---Finally we demonstrate our theory in a lattice model with broken time-reversal symmetry, which can be tuned from topological trivial to nontrivial.  We show that even though our theory is a bulk one, it can capture the contribution due to the chiral edge states.

Consider the following Hamiltonian~\cite{qi2006},
\begin{equation}\label{insulatorH}
H(\bm k)=\sum_{i=1,2}\left(\sigma_{i}\sin k_{i}-\sigma_{3}\cos k_i\right)+\lambda\sigma_{3} \;,
\end{equation}
where $\sigma_{i=1,2,3}$ are the Pauli matrices.  We assume that the Fermi energy lies in the gap.  When $-2 < \lambda < 0$ or $0 < \lambda < 2$, the system is a Chern insulator with the Chern number $C=-1$ or $C=1$, respectively; otherwise it is a topologically trivial insulator with $C = 0$.  Figure~\ref{fig:MQ}(a) shows the $\mu$-dependence of $\bm M_Q$ for $C = 1$ and $C = 0$.  As we can see, $\bm M_Q$ stays almost constant inside the gap when $C = 0$ while it shows significant change when $C = 1$.  This is due to the existence of the chiral edge states in the latter.  The $T$-dependence of $\bm M_Q$ shows similar behavior [Fig.~\ref{fig:MQ}(b)].

\begin{figure}[t]
  \includegraphics[width=1.0\columnwidth]{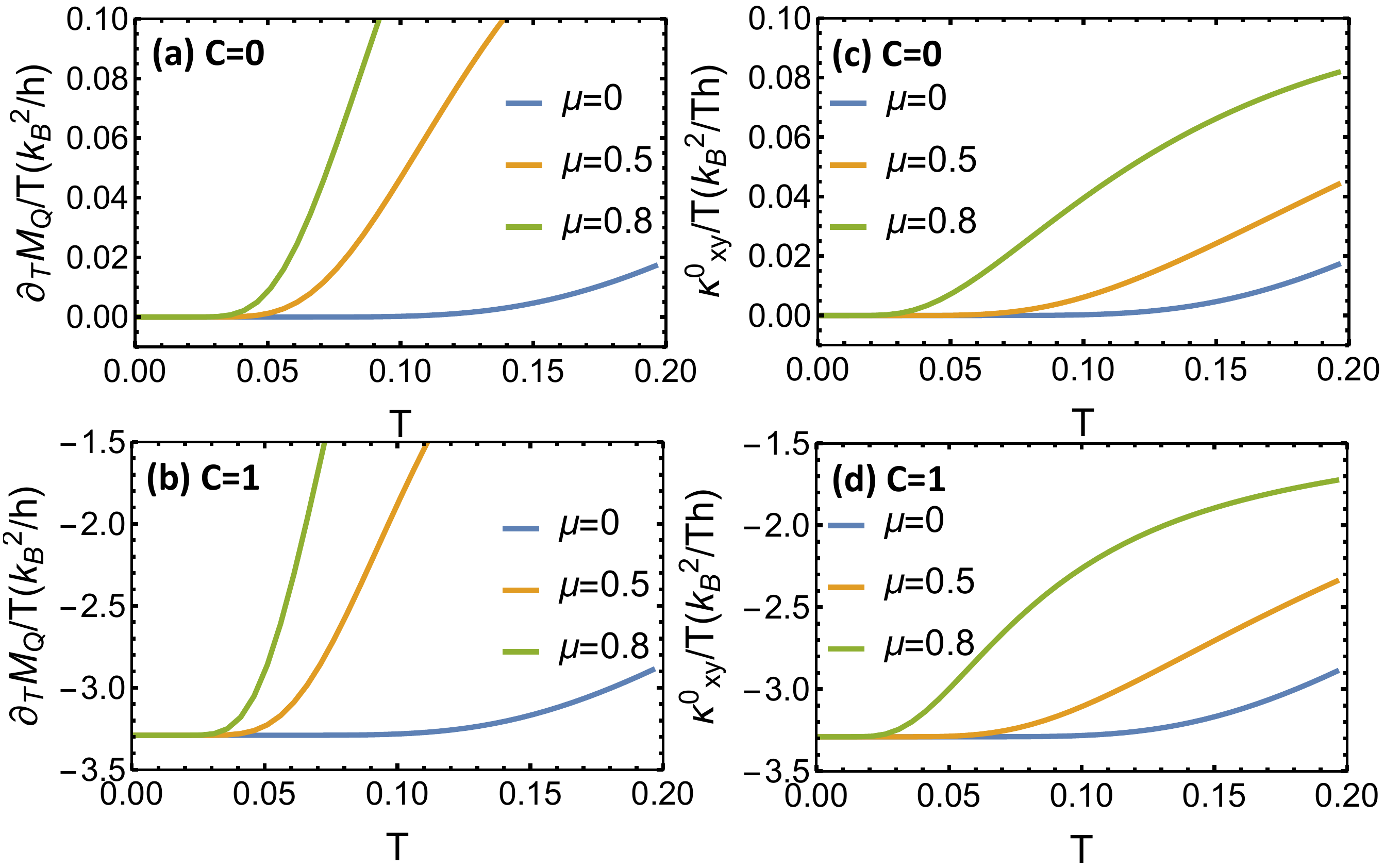} 
  \caption{The temperature dependence of thermal Hall conductivity in the Chern insulator from the Streda formula $\partial_{T}\bm M_{Q}/T$ in (a)(b), and from the linear response theory $\kappa^0_{xy}/T$ in (c)(d). We choose $\lambda=-3$ for the cases of $C=0$, and $\lambda=-1$ for those of $C=1$. }
\label{fig:streda}
\end{figure}

Next we plot the thermal Hall conductivity using the Streda formula in Fig.~\ref{fig:streda}.  When $C = 0$, $\kappa_{xy}/T \to 0$ as $T \to 0$, and when $C = 1$, $\kappa_{xy}/T \to \pi^2k_B^2/3h$.  This is the expected behavior from the Wiedemann-Franz law. The Streda formula also contains a contribution from the heat magnetic moment $\bm m^Q$ of the wave packet, which should vanish in the low-temperature limit. In Fig.~\ref{fig:streda}(c) and (d), we plot the thermal Hall conductivity from the linear response calculation~\cite{qin2011}, which does not include $\bm m^Q$.  We can see that the Streda formula and the linear response calculation agrees well at low temperatures.  


In summary, we have constructed the thermodynamics of the heat magnetization in the presence of the gravitomagnetic field.  Based on the modified free energy, we derived the explicit expression of the heat magnetization by static response theory in equilibrium, and rigorously established the Streda formula for the thermal Hall effect.  The modified free energy implies that the entropy is also modified by the gravitational magnetic field $\bm B_g$ and it will be interesting to explore its consequences.

We thank Zhengqian Cheng, Qian Niu, Ying Ran, Junren Shi, Qin Tao, Chong Wang, and Cong Xiao for stimulating discussions.  This work was supported by the Department of Energy, Basic Energy Sciences, Materials Sciences and Engineering Division, Pro-QM EFRC (DE-SC0019443). D.X. also acknowledges the support of a Simons Foundation Fellowship in Theoretical Physics.

%

\newpage

\vskip 15 cm
\begin{widetext}
\setcounter{figure}{0}
\setcounter{equation}{0}
\setcounter{section}{0}

\renewcommand\thefigure{S\arabic{figure}}
\renewcommand\theequation{S\arabic{equation}}

\section*{Supplementary Information: Thermodynamics of Energy magnetization}
\subsection{Calculation of the heat magnetization using the Maxwell relation Eq.~(15)}

In this section, we show how to calculate heat magnetization using the Maxwell relation, Eq.~(15) in the main text, which we reproduce below
\begin{equation}
\Bigl(\frac{\partial\bm M_Q}{\partial\mu}\Bigr)_{\bm B_g, T} = 
\Bigl(\frac{\partial N}{\partial\bm B_g}\Bigr)_{T, \mu} \;.
\end{equation}

Our starting point is the static response function between the particle number density and heat current, given by 
\begin{equation}\label{responsef_unction}
\bm\Pi_{n,j^{Q}}(\bm q)=\sum_{\alpha \beta} n_{\alpha\beta}({\bm q})\bm j^{Q}_{\beta\alpha}(-\bm q)\frac{f_\alpha-f_\beta}{\varepsilon_{\alpha}-\varepsilon_{\beta}} \;,
\end{equation}
where $n_{\alpha\beta}(\bm q)=\langle \psi_{\alpha}|\hat{\bm n}_{\bm q}|\psi_{\beta}\rangle$ and $\bm j^{Q}_{\beta\alpha}(-\bm q)=\langle \psi_{\beta}|\hat{\bm j}^{Q}_{-\bm q}|\psi_{\alpha}\rangle$ are the matrix elements of the particle number density operator and heat current density operator, respectively, and $f_{\alpha}$ is the Fermi-Dirac distribution for the single-particle energy $\varepsilon_{\alpha}$.  In a periodic system, the matrix element of the density operator is 
\begin{equation}\label{density}
\langle \psi_{n\bm k}|\hat{n}(\bm q)|\psi_{m\bm k^\prime}\rangle =\langle \psi_{n\bm k}|e^{-i\bm q\cdot \hat{\bm r}}|\psi_{m\bm k^\prime}\rangle =\langle u_{n\bm k}|u_{m\bm k+\bm q}\rangle\delta_{\bm k^\prime,\bm k+\bm q} \;,
\end{equation}
where in the last step we have used the translation invariance in a solid. Similarly, the matrix element of the heat current operator reads
\begin{equation}\label{heat_current}
\langle \psi_{n\bm k}|\hat{\bm j}^{Q}(\bm q)|\psi_{n^\prime\bm k^{\prime}}\rangle=\langle u_{n\bm k}|\frac{\{\hat{\bm v}_{\bm k},\hat{\mathcal{K}}_{\bm k}\}+\{\hat{\bm v}_{\bm k+\bm q},\hat{\mathcal{K}}_{\bm k+\bm q}\}}{4}|u_{n^\prime,\bm k^\prime}\rangle\delta_{\bm k^{\prime}, \bm k+\bm q} \;,
\end{equation} 
where $\hat{\bm v}_{\bm k}\equiv e^{-i\bm k\cdot\hat{\bm r}}\hat{\bm v} e^{i\bm k\cdot\hat{\bm r}}$ is the velocity operator, and $\hat{\mathcal{K}}_{\bm k}\equiv e^{-i\bm k\cdot \hat{\bm r}}\hat{\mathcal{K}} e^{i\bm k\cdot\hat{\bm r}}$ is the single-particle Hamiltonian operator; both in the Bloch basis.  $|u_{n\bm k}\rangle$ is the periodic part of the Bloch wave function $|\psi_{n\bm k}\rangle$.  Inserting Eq.~\eqref{density} and Eq.~\eqref{heat_current} into Eq.~\eqref{responsef_unction}, we obtain the explicit form of the response function $\bm\Pi_{n,j^{Q}}(\bm q)$, 
\begin{equation}
\bm\Pi_{n,j^{Q}}(\bm q)=\sum_{n,m,\bm k}\langle u_{n\bm k}|u_{m\bm k+\bm q}\rangle \frac{f_{n\bm k}-f_{m\bm k+\bm q}}{\varepsilon_{n\bm k}-\varepsilon_{m\bm k+\bm q}}\langle u_{m\bm{k}+\bm{q}}|\frac{\{\hat{\mathcal{K}}_{\bm{k}+\bm{q}},\hat{\bm{v}}_{\bm{k}+\bm q}\}+\{\hat{\bm{v}}_{\bm{k}},\hat{\mathcal{K}}_{\bm{k}}\}}{4}|u_{n\bm{k}}\rangle \;.
\end{equation}

The density induced by the gravitational vector potential $\bm A_g(\bm q)$ is given by $\delta n_{\bm q} = \bm\Pi_{nj^Q(\bm q)} \cdot \bm A_g(\bm q)$.  To extract the response to $\bm B_g = \nabla \times \bm A_g$, we expand $\bm\Pi_{nj^Q(\bm q)}$ up to first order in $\bm q$ and only retain the antisymmetric part after taking the limit $\bm q \to 0$.  To this end, we introduce the auxiliary heat magnetization,   
\begin{equation}\label{auxilaryM}
\tilde{\bm M}_{Q}\equiv\lim_{\bm q\rightarrow 0}\frac{i}{2}\nabla_{\bm q}\times \bm{\Pi}_{nj}(\bm q) \;.
\end{equation}
We first consider the interband contribution of $\tilde{\bm M}_{Q}$ when $n\neq m$ at $\bm q\rightarrow 0$, which is given by
\begin{align}
\tilde{M}_{Qi}^{inter}&=-\lim_{\bm{q}\rightarrow0}\frac{i}{2}\epsilon_{ijk}\partial_{j}\sum_{n\neq m}\sum_{\bm{k}}\langle u_{n\bm{k}}|\frac{\hat{\mathcal{K}}_{\bm{k}}+\hat{\mathcal{K}}_{\bm{k}+\bm{q}}}{2}|u_{m\bm{k}+\bm{q}}\rangle\langle u_{m\bm{k}+\bm{q}}|\frac{v_{\bm{k}}^{k}\mathcal{K}_{\bm{k}}+\mathcal{K}_{\bm{k}+\bm{q}}v_{\bm{k}+\bm{q}}^{k}}{2}|u_{n\bm{k}}\rangle\frac{f_{n\bm{k}}-f_{m\bm{k}+\bm{q}}}{\varepsilon_{n\bm{k}}-\varepsilon_{m\bm{k}+\bm{q}}},\\&=\frac{i}{4}\epsilon_{ijk}\sum_{n\neq m}\sum_{\bm{k}}\left(A_{\bm{k}nm}^{j}(\tilde{\varepsilon}_{m\bm{k}}+\tilde{\varepsilon}_{n\bm{k}})^{2}A_{\bm{k}mn}^{k}\right)f_{n\bm{k}}\\
&=-\sum_{n\bm{k}}f_{n\bm{k}}(\bm m_{n\bm{k}}+\tilde{\varepsilon}_{n}\bm{\Omega}_{n\bm{k}}) \;,
\end{align}
where $f_{n\bm k}=f(\tilde{\varepsilon}_{n\bm k})$, $\bm m_{n{\bm k}}$ is the orbital magnetic orbital momentum for the $n$th band~\cite{xiao2006}
\begin{equation}
\bm m_{n\bm k}\equiv\frac{i}{2}\sum_{m\neq n}(\varepsilon_{m\bm k}-\varepsilon_{n\bm k})\bm{A}_{\bm k nm}\times\bm{A}_{\bm k mn},
\end{equation}
and $\bm\Omega_{nk} = \nabla_{\bm k} \times \bm A_{\bm knn}$ is the Berry curvature with $\bm A_{\bm k nm}\equiv \langle u_{n\bm k}|i\partial_{\bm k}|u_{m\bm k}\rangle$ the Berry connection. 

The intraband contribution $\tilde{\bm M}^{intra}_{Q}$ at $\bm q\rightarrow 0$ when $n=m$ is given by
\begin{align}
\tilde{M}^{intra}_{Q i}=&\lim_{\bm q\rightarrow 0}\frac{i}{2}\epsilon_{ijk}\partial_{q_{j}}\sum_{n,\bm{k}}\frac{f_{n\bm{k}}-f_{n\bm{k}+\bm{q}}}{\varepsilon_{n\bm{k}}-\varepsilon_{n\bm{k+\bm{q}}}}\langle u_{n\bm{k}+\bm{q}}|\frac{\{\hat{\mathcal{K}}_{\bm{k}+\bm{q}},\hat{v}_{\bm{k}+\bm{q}}^{k}\}+\{\hat{v}_{\bm{k}}^{k},\hat{\mathcal{K}}_{\bm{k}}\}}{4}|u_{n\bm{k}}\rangle \langle u_{n\bm k}|u_{n\bm{k}+\bm{q}}\rangle\\
=&\frac{i}{2}\epsilon_{ijk}\sum_{n,\bm{k}}\langle u_{n}|\partial_{q_{j}}u_{n\bm{k}}\rangle\tilde{\varepsilon}_{n\bm{k}}v_{\bm{k}nn}^{k}f_{n\bm{k}}^{\prime}+\frac{i}{2}\varepsilon_{ijk}\sum_{n,\bm{k}}\left(\frac{\tilde{\varepsilon}_{m\bm{k}}+\tilde{\varepsilon}_{n\bm{k}}}{2}\langle\partial_{j}u_{n\bm{k}}|u_{m\bm{k}}\rangle v_{\bm{k}mn}^{k}\right)f_{n\bm{k}}^{\prime}\\
=&-\frac{i}{4}\epsilon_{ijk}\sum_{n\neq m,\bm{k}}A_{\bm{k}nm}^{j}A_{\bm{k}mn}^{k}(\tilde{\varepsilon}_{m\bm{k}}^{2}-\tilde{\varepsilon}_{n\bm{k}}^{2})f_{n\bm{k}}^{\prime}\\
=&-\sum_{n\bm k}\bm{m}^Q_{n\bm k}f_{n\bm{k}}^{\prime} \;,
\end{align}
where $v^k_{\bm k nm}=\langle u_{n\bm k}|\hat{v}^k_{\bm k}|u_{m\bm k}\rangle$ and the heat magnetic momentum $\bm m_{n\bm k}^Q$ is derived in Sec.~\ref{orbital_heatmag}. Collecting both the intraband and interband contributions, the auxiliary heat magnetization is 
\begin{align}
\tilde{\bm{M}}_{Q}=&-\sum_{n\bm{k}}\left(f_{n\bm{k}}(\bm{m}_{n\bm{k}}+\tilde{\varepsilon}_{n\bm{k}}\bm{\Omega}_{n\bm{k}})-\tilde{\varepsilon}_{n\bm{k}}\bm{m}^{Q}_{n\bm k}f_{n\bm{k}}^{\prime}\right).
\end{align}
Integrating the auxiliary heat magnetization with respect to the chemical potential and using the boundary condition $\bm M_{Q}\rightarrow 0$ at $\mu\rightarrow-\infty$, we can obtain the heat magnetization in the main text. 

\subsection{Calculation of the heat magnetization using the Maxwell relation Eq. (16)}

In this section, we calculate the heat magnetization using the Maxwell relation Eq.~(16) in the main text,
\begin{equation} \label{entropyy}
\Bigl(\frac{\partial\bm M_Q}{\partial T}\Bigr)_{\bm B_g, \mu} = 
\Bigl(\frac{\partial S}{\partial\bm B_g}\Bigr)_{T, \mu} \;.
\end{equation}

Let us first establish the relation between the change of heat magnetization with respect to the temperature and the response function $\bm{\Pi}_{\mathcal{K},j^{Q}}(\bm{q})$ between $\hat{\mathcal K}_{\bm q}$ and $\bm j^{Q}_{-\bm q}$. According to the modified free energy, ones can verify the thermodynamical relation
\begin{equation}\label{thermodynamics}
K=\Omega_{b} + TS - \partial_{\beta}(\beta\delta \Omega_{b}) \;,
\end{equation}
where $K=\int d\bm r \text{Tr}[\hat{\rho}_{A}\hat{\mathcal{H}}_{A}]$, and $\delta \Omega_{b}=-\int d\bm r \bm B_g\cdot\bm M_{N}$ is the modification to $\Omega_{b}$. Differentiating Eq.~\eqref{thermodynamics} with respect to $\bm B_{g}$ at fixed $T$ and $\mu$, we have 
\begin{equation}
-\frac{\partial{K}}{\partial{\bm B_{g}}}=\bm M_{Q}-T\frac{\partial S}{\partial \bm B_{g}}-\frac{\partial (\beta\delta \bm M_{Q})}{\partial\beta}
\end{equation}   
with $\delta \bm M_{Q}=-\partial\delta \bm \Omega_{b}/\partial\bm B_{Q}$. Inserting the Maxwell relation~\eqref{entropyy} into the above equation, we obtain
\begin{equation}\label{entropy}
-\frac{\partial{K}}{\partial{\bm B_{Q}}}=\frac{\partial (\beta \bm M_{Q}-\beta \delta \bm M_{Q})}{\partial \beta} \;.
\end{equation}

We can then calculate the heat magnetization following the same steps as in the previous section.  We first introduce the auxiliary heat magnetization $\tilde{\bm M}_{Q}\equiv-\frac{\partial{K}}{\partial{\bm B_{Q}}}$, which can be related to the response function
\begin{equation}
\bm{\Pi}_{\mathcal{K},j^{Q}}(\bm{q})=-\lim_{\omega_n\rightarrow 0}\int d\tau e^{i\tau \omega_{n}}\langle T\hat{\rho}_{A=0} \hat{\mathcal{K}}_{\bm q}(\tau)\bm{\hat{j}}^{Q}_{-\bm q}(0) \rangle
\end{equation}  
via the following equation
\begin{align}
\tilde{\bm M}_{Q}\equiv-\frac{i}{2}\lim_{\bm{q}\rightarrow0}\nabla_{\bm{q}}\times\bm{\Pi}_{\mathcal{K},j^{Q}}(\bm{q}) \;.
\end{align}

In Bloch basis, the response function $\bm{\Pi}_{\mathcal{K},j^{Q}}(\bm{q})$ is given by
\begin{align}
&\bm{\Pi}_{\mathcal{K},j^{Q}}(\bm{q})=\sum_{\bm{k}mn}\langle u_{n\bm{k}}|\frac{\hat{\mathcal{K}}_{\bm{k}}+\hat{\mathcal{K}}_{\bm{k}+\bm{q}}}{2}|u_{m\bm{k}+\bm{q}}\rangle\langle u_{m\bm{k}+\bm{q}}|\frac{\{v_{\bm{k}+\bm{q}},\mathcal{K}_{\bm{k}+\bm{q}}\}+\{\mathcal{K}_{\bm{k}}, v_{\bm{k}}\}}{4}|u_{n\bm{k}}\rangle\frac{f_{n\bm{k}}-f_{m\bm{k}+\bm{q}}}{\varepsilon_{n\bm{k}}-\varepsilon_{m\bm{k}+\bm q}} \;.
\end{align}
The the inerband contribution to $\bm M^{au}_{Q}$ at $\bm q \to 0$ reads
\begin{align}
\tilde{M}^{inter}_{Q i}&=-\lim_{\bm{q}\rightarrow0}\frac{i}{2}\epsilon_{ijk}\partial_{q_j}\sum_{\bm k, n\neq m}\langle u_{n\bm{k}}|\frac{\hat{\mathcal{K}}_{\bm{k}}+\hat{\mathcal{K}}_{\bm{k}+\bm{q}}}{2}|u_{m\bm{k}+\bm{q}}\rangle
\langle u_{m\bm{k}+\bm{q}}|\frac{\{\hat{v}_{\bm{k}+\bm{q}}^{k},\hat{\mathcal{K}}_{\bm{k}+\bm q}\}+\{\hat{\mathcal{K}}_{\bm{k}},\hat{v}_{\bm{k}}^{k}\}}{4}|u_{n\bm{k}}\rangle
\frac{f_{n\bm{k}}-f_{m\bm{k}+\bm{q}}}{\varepsilon_{n\bm{k}}-\varepsilon_{m\bm{k}+\bm{q}}}\\
&=\frac{i}{4}\epsilon_{ijk}\sum_{\bm k,n\neq m}\left(A_{\bm{k}nm}^{j}(\tilde{\varepsilon}_{m\bm{k}}+\tilde{\varepsilon}_{n\bm{k}})^{2}A_{\bm{k}mn}^{k}\right)f_{n\bm{k}}\\
&=\frac{i}{4}\sum_{\bm k,n\neq m}\left(\bm A_{\bm{k}nm}\times(\tilde{\varepsilon}_{m\bm{k}}+\tilde{\varepsilon}_{n\bm{k}})^{2}\bm A_{\bm{k}mn}\right)f_{n\bm{k}} \;,
\end{align}  
where we have used $\hat{\mathcal K}_{\bm q}|u_{n\bm q}\rangle=\tilde{\varepsilon}_{n\bm q}|u_{n\bm q}\rangle$.  The intraband contribution to $\tilde{\bm M}_{Q}$ at $\bm q\rightarrow 0$ is
\begin{align}
\tilde{M}^{intra}_{Q i}=&-\lim_{\bm q\rightarrow 0}\frac{i}{2}\epsilon_{ijk}\partial_{q_j}\sum_{\bm{k},n}\langle u_{n\bm{k}}|\frac{\hat{\mathcal{K}}_{\bm{k}}+\hat{\mathcal{K}}_{\bm{k}+\bm{q}}}{2}|u_{n\bm{k}+\bm{q}}\rangle\langle u_{n\bm{k}+\bm{q}}|\frac{\{\hat{v}_{\bm{k}+\bm{q}}^{k},\hat{\mathcal{K}}_{\bm{k}+\bm q}\}+\{\hat{\mathcal{K}}_{\bm{k}},\hat{v}_{\bm{k}}^{k}\}}{4}|u_{n\bm{k}}\rangle\frac{f_{n\bm{k}}-f_{n\bm{k}+\bm{q}}}{\varepsilon_{n\bm{k}}-\varepsilon_{n\bm{k}+\bm{q}}}\\
=&\epsilon_{ijk}\sum_{\bm{k},n}\left(-\frac{i}{4}v_{\bm{k}nn}^{j}v_{\bm{k}nn}^{k}\tilde{\varepsilon}_{n\bm{k}}f_{n\bm{k}}^{\prime}-\frac{1}{2}\tilde{\varepsilon}^2_{n\bm{k}}A_{nn}^{j}v_{\bm{k}nn}^{k}f_{n\bm{k}}^{\prime}-\frac{i}{4}\tilde{\varepsilon}_{n\bm{k}}\langle\partial_{j}u_{n\bm{k}}|v_{\bm{k}}^{k}\mathcal{K}_{\bm{k}}+\mathcal{K}_{\bm{k}}v_{\bm{k}}^{k}|u_{n\bm{k}}\rangle f_{n\bm{k}}^{\prime}\right)\\
=&-\frac{i}{4}\epsilon_{ijk}\sum_{n\bm k}\tilde{\varepsilon}_{n\bm{k}}f_{n\bm{k}}^{\prime}\left(\sum_{m\neq n}iA_{\bm{k}nm}^{j}v_{\bm{k}mn}^{k}(\tilde{\varepsilon}_{n\bm{k}}+\tilde{\varepsilon}_{m\bm{k}})\right)\\
=&\sum_{n\bm k}\tilde{\varepsilon}_{n\bm{k}}f_{n\bm{k}}^{\prime}\bm m^{Q}_{n\bm{k}} \;.
\end{align}
Collecting both the intra- and inter-band contribution, we find the auxiliary heat magnetization is given by
\begin{align}
\tilde{\bm M}_{Q}=\frac{i}{4}\sum_{n\neq m}\sum_{\bm{k}}\bm A_{\bm{k}nm}\times \bm A_{\bm{k}mn}(\tilde{\varepsilon}_{m\bm{k}}+\tilde{\varepsilon}_{n\bm{k}})^{2}f_{n\bm{k}}+\sum_{n\bm k}\tilde{\varepsilon}_{n\bm{k}}f_{n\bm{k}}^{\prime}\bm m^{Q}_{n\bm{k}} \;.
\end{align}

Finally, following Eq.~\eqref{entropy}, we integrate $\tilde{\bm M}_{Q}$ with respect to $\beta$ from $\infty$ to $\beta$ and obtain
\begin{align}
\bm M_{Q}&=\frac{i}{4}\sum_{n\neq m}\sum_{\bm{k}}\bm A_{\bm{k}nm}\times \bm A_{\bm{k}mn}(\tilde{\varepsilon}_{m\bm{k}}+\tilde{\varepsilon}_{n\bm{k}})^{2}\frac{g_{n\bm{k}}}{\varepsilon_{n\bm{k}}}\\&+\sum_{n\bm k} \left(\bm m^{Q}_{n\bm{k}}\frac{\tilde{\varepsilon}_{n\bm{k}}f_{n\bm{k}}-g_{n\bm{k}}}{\tilde{\varepsilon}_{n\bm{k}}}-\bm m_{n\bm{k}}g_{n\bm{k}}-\int^{\tilde{\varepsilon}_{n\bm{k}}}_{\infty}d\lambda \bm \Omega_{n\bm{k}}g(\lambda)\right)\\&=\sum_{n\bm k}\left(\int_{\infty}^{\tilde{\varepsilon}_{n\bm{k}}}\lambda d\lambda\bm\Omega_{n}(\bm k)f(\lambda)+\bm m^Q_{n\bm{k}}f_{n\bm{k}}\right),
\end{align} 
where $g(x)=-\ln(1+e^{-\beta x})/\beta$ and $g_{n\bm k}=g(\tilde{\varepsilon}_{n\bm k})$, and we have used the following identities, 
\begin{align}
\int_{\infty}^{\beta}f(\varepsilon)d\lambda&=\frac{\beta}{\varepsilon}g(\varepsilon) \;,\\
\int_{\infty}^{\beta}f^{\prime}(\varepsilon)d\lambda&=\frac{1}{\varepsilon}\beta f(\varepsilon)-\frac{\beta}{\varepsilon^{2}}\int_{\infty}^{\epsilon}f(\lambda)d\lambda \;.
\end{align}

\subsection{The heat orbital magnetic momentum}\label{orbital_heatmag}

Consider a wave packet composed of Bloch wave functions $|\psi_{n\bm k}\rangle$ from the $n$th band,
\begin{equation}\label{wave-packet}
|W\rangle=\int d\bm k a(\bm k)|\psi_{n\bm k}\rangle \;.
\end{equation}
We assume that the wave packet is centered around $\bm k_c$ and $\bm r_c$ in the phase space, i.e., $|a(\bm k, t)|^2 \approx \delta(\bm k-\bm k_{c})$ and 
\begin{equation}
\bm r_{c}\equiv\langle W|\hat{\bm r}|W\rangle=\partial_{\bm k_{c}}\gamma(\bm k_{c},t)+\bm A_{\bm k_{c}nn},
\end{equation}
with $\gamma(\bm k, t)$ being the phase factor $a(\bm k, t)=|a(\bm k, t)|e^{-i\gamma(\bm k,t)}$~\cite{sundaram1999}.

The operator of the heat orbital momentum is defined as 
\begin{equation}
\hat{\bm{m}}^{Q}=\frac{1}{4}\left(\hat{\bm{r}}-\bm{r}_{c}\right)\times\hat{\bm{j}}_{Q}+ \text{h.c}.
\end{equation}
Its expectation value with respect to the wave packet is given by
\begin{align}
&\langle W| \hat{\bm m}_{Q}|W\rangle=\frac{1}{4}\langle W|\delta\hat{\bm{r}}\times\frac{1}{2}\{\hat{\mathcal{K}},\hat{\bm{v}}\}|W\rangle+ \text{c.c}\\
=&\frac{1}{4}\epsilon_{ijk}\int d\bm{k}d\bm{k}^{\prime}a^{\ast}(\bm{k},t)a(\bm{k}^{\prime},t)\langle\psi_{n\bm{k}}|\delta\hat{r}_{j}|\psi_{m\bm{p}}\rangle\langle\psi_{m\bm{p}}|\frac{\{\hat{\mathcal{K}},\hat{\bm{v}}\}}{2}|\psi_{n\bm{k}^{\prime}}\rangle+ \text{c.c}\\
=&\frac{1}{4}\epsilon_{ijk}\int d\bm{k}d\bm{k}^{\prime}a^{\ast}(\bm{k},t)a(\bm{k}^{\prime},t)(i\partial_{k_{j}}\delta_{nm}+A_{\bm{k}nm}^{j}-r_{c}^{j}\delta_{nm})\delta(\bm{k}-\bm{k}^{\prime})\frac{\tilde{\varepsilon}_{m\bm{k}^{\prime}}+\tilde{\varepsilon}_{n\bm{k}^{\prime}}}{2}v_{mn\bm{k}^{\prime}}^{k}+ \text{c.c}\\
=&\frac{1}{4}\epsilon_{ijk}\int d\bm{k}(-i\partial_{\bm{k}}a^{\ast}(\bm{k},t))a(\bm{k},t)\tilde{\varepsilon}_{n\bm{k}}v_{nn\bm{k}}^{k}
+ \frac{1}{4}\epsilon_{ijk}(A^j_{\bm{k}_{c}nm}-r^j_{c}\delta_{mn})\frac{\tilde{\varepsilon}_{m\bm{k}_{c}}+\tilde{\varepsilon}_{n\bm{k}_{c}}}{2}v_{mn\bm{k}_{c}}^{k}+c.c\\
=&\frac{1}{4}\epsilon_{ijk}A^j_{\bm k_{c}nm}\frac{\tilde{\varepsilon}_{m\bm{k}_{c}}+\tilde{\varepsilon}_{n\bm{k}_{c}}}{2}v_{mn\bm{k}_{c}}^{k}+c.c\\
=&\bm m^{Q}_{n\bm k_{c}},
\end{align}
where $\delta\hat{\bm r}\equiv\hat{\bm r}-\bm r_{c}$ and we have used the completeness relation, 
\begin{equation}
\sum_{n}\int d\bm k |\psi_{n\bm k}\rangle\langle \psi_{n \bm k}|=1 \;.
\end{equation}

\subsection{Thermal Hall conductivity}\label{Appendix:coarse-graining}

Using the heat magnetization we can derive the thermal Hall conductivity using the semiclassical theory developed in Ref.~\cite{xiao2006}  The advantage of this method is that it can deal with statistical forces directly without introducing their mechanical counterparts, and need no special gauge choice for the heat current density operator. 


In the presence of an inhomogeneous temperature distribution $T(\bm r)$, the local heat current can be calculated using the semiclassical coarse graining method,
\begin{equation}\label{coarsegraining}
\bm J_{Q}(\bm r)=\frac{1}{4}\int [d\bm q_{c}]d\bm r_{c}\left[\langle W|\{\hat{\bm v},\hat{\mathcal K}\}\delta(\bm{r}-\hat{\bm{r}})|W\rangle\right]+ \text{h.c.}
\end{equation}   
where the wave-packet $|W\rangle$ is centered at $\{\bm r_{c},\bm q_{c}\}$, and $\hat{\mathcal{K}}$ is the single-particle Hamiltonian. The coarse graining is implemented by expanding $\delta(\bm r-\hat{\bm r})$ around $\bm r-\bm r_{c}$
\begin{equation}\label{delta}
\delta(\bm r -\hat{\bm r}) = \delta(\bm r - \bm r_{c})-(\hat{\bm r}-\bm r_{c})\cdot \nabla\delta(\bm r -\bm r_{c})+ O(\hat{\bm r}-\bm r_{c}) \;.
\end{equation}
Inserting Eq.~\eqref{delta} into Eq.~\eqref{coarsegraining}, we obtain
\begin{equation}\label{JQQ}
\begin{split}
\bm J_{Q}(\bm r)=\int [d\bm{q}]\left(f(\bm{q},\bm{r})\tilde{\varepsilon}_{\bm{q}}\bm{v}_{\bm{q}},
+\nabla\times f(\bm{q},\bm{r})\bm{m}^{Q}_{\bm q}\right) \;,
\end{split}
\end{equation} 
where $\bm{v}_{\bm{q}}\equiv\partial_{\bm q} \tilde{\varepsilon}_{\bm q}$ is the band velocity, and the Fermi-Dirac distribution $f(\bm q, \bm r)$ is defined with respect to the local temperature $T(\bm r)$.  For simplicity, summation over the band index is omitted. The first term in Eq.~\eqref{JQQ} vanishes because the integrand can be written as a total derivative with respect to $\bm q$.  We are thus left with the heat magnetic moment $\bm m_{\bm q}^Q$ only.

The local heat current includes both the transport current and the magnetization current.  The latter needs to be subtracted~\cite{xiao2006,cooper1997}.  The transport current is given by
\begin{equation}\label{thermalcurrent}
\bm J^{tr}_{Q}(\bm r)= \bm J_Q - \nabla \times \bm M_Q = -\frac{\bm{\nabla} T}{T}\times\int[d\bm{q}]\int^{\infty}_{\tilde{\varepsilon}_{\bm{k}}}d\lambda\partial_{\lambda}f(\lambda)\lambda^{2}\bm{\Omega}_{\bm q} \;.
\end{equation} 
We can extract the thermal Hall conductivity $\kappa^0_{xy}$, defined by $\bm J^{tr}_{Q x}=\kappa^0_{xy}(-\nabla_y T)$, from the above equation.  After some algebra, we find
\begin{equation}
\kappa^0_{xy}=(-T)\int[d\bm{q}]c(f(\tilde{\varepsilon}_{\bm{q}}))\Omega_{z\bm q} \;.
\end{equation}
where the weight function for fermion Dirac distribution $\rho$ is $c(\rho)=(\rho-1)\ln^{2}(\rho^{-1}-1)+\ln^{2}\rho+2\text{Li}_{2}(\rho)$ with $\text{Li}_2(x)$ being polylogarithm function. Note the coefficient $\kappa^0_{xy}$ has the unit $k^2_{B}/\hbar$.  The calculated $\kappa^0_{xy}$ using the heat magnetization is the same as that using the linear response theory~\cite{qin2011}.

\end{widetext}

\end{document}